УДК 378.091:004.4

**Величко Владислав Євгенович**
канд. фіз-мат наук, доцент, докторант
ДЗ «Луганський національний університет імені Тараса Шевченка», м. Старобільськ, Україна
*vladislav.velichko@gmail.com*

# СТРАТЕГІЯ ВПРОВАДЖЕННЯ ВІЛЬНОГО ПРОГРАМНОГО ЗАБЕЗПЕЧЕННЯ В ПРОЦЕС ПІДГОТОВКИ ВЧИТЕЛІВ МАТЕМАТИКИ, ФІЗИКИ ТА ІНФОРМАТИКИ

**Анотація.** Інформаційні процеси в суспільстві спонукають освіту до перегляду форм і методів навчання, яке передбачає використання широких дидактичних можливостей інформаційно-комунікаційних технологій. Не менш значущою в цьому контексті є проблема підготовки фахівців, які здатні використовувати сучасні можливості інформаційних технологій. Підготовка висококваліфікованого вчителя можлива тільки за умови використання в його навчанні передових технологій, які охоплюють весь спектр існуючих можливостей. Аналіз програмного забезпечення супроводу навчальної діяльності студентів вишів показав недостатнє використання в освітньому процесі цілого класу програмного забезпечення – вільного програмного забезпечення. Для подолання цієї проблеми запропонована стратегія впровадження вільного програмного забезпечення у підготовці вчителів математики, фізики та інформатики.

**Ключові слова:** вільне програмне забезпечення; стратегія впровадження; підготовка вчителів математики, фізики та інформатики.

## 1. ВСТУП

**Постановка проблеми.** Інформаційно-комунікаційні технології в сучасному інформаційному суспільстві, з одного боку, є ефективним інструментом підвищення якості освіти, спираючись на інтерактивність, підвищення ролі самостійності, використання мультимедійних і комунікаційних можливостей, а, з іншого, є повноцінним середовищем розвитку соціального суб'єкта суспільства. Однією із задач сучасної вищої педагогічної освіти є підготовка і перепідготовка фахівців у галузі освіти, які не тільки вміють використовувати інформаційно-комунікаційні технології в своїй професійній діяльності, а й досконало володіють способами й особливостями їх використання у своїй галузі і за її межами [10].

Якщо для майбутніх фахівців з гуманітарних напрямків інформаційно-комунікаційні технології переважно є інструментом навчання, то для майбутніх учителів математики, фізики та інформатики інформаційно-комунікаційні технології є ще й предметом детального вивчення і постійного використання. Враховуючи це, необхідно організувати інформатичну підготовку вчителів математики, фізики та інформатики так, щоб вона не була прив'язана тільки до певного програмного забезпечення, або принаймні, надавала всебічні знання з усіх можливих розділів, базуючись на різноманітному програмному забезпеченні.

Аналіз існуючих навчальних планів підготовки вчителів математики, фізики та інформатики, а також робочих програм з дисциплін інформатичної підготовки даних напрямів свідчить про те, що великий сегмент програмного забезпечення, який називають вільним програмним забезпеченням, повноцінно не використовується, і навіть не розглядається. Це протиречить задачам всебічної підготовки висококваліфікованих фахівців та ідеалам вільної освіти взагалі. Отже, проблема впровадження вільного програмного забезпечення в процесі підготовки майбутніх учителів математики, фізики та інформатики потребує нагального розв'язання.

**Аналіз останніх досліджень і публікацій.** Загальні проблеми вільного програмного забезпечення, юридичні та філософські аспекти його існування та використання висвітлюються в роботах Дж. Гослінга, Е. Реймонда, Р. Столлмана та ін. Окремим аспектам використання вільного програмного забезпечення в системі освіти присвятили свої роботи Є. Алексєєв, В. Габрусєв, О. Дима, Г. Злобін, М. Карпенко, М. Кияк, О. Нестеренко, Л. Панченко, А. Панчук, С. Семеріков, І. Теплицький, В. Хахановський та ін. [2;3; 4; 5; 6; 7; 11; 12]. Але проблеми стратегії впровадження вільного програмного забезпечення в процес підготовки майбутніх учителів фізики, математики та інформатики осталась поза увагою дослідників.

Сучасні уявлення щодо якісної освіти мають декілька аспектів. По-перше, якість освіти залежить від того, якою мірою вона задовольняє потреби суб'єктів освітнього процесу, надаючи їм можливість формувати власну освітню траєкторію. По-друге, якість освіти передбачає формування нової системи універсальних базових компетентностей, у тому числі й самоосвітньої компетентності. По-третє, важним аспектом підвищення якості освіти є широке використання інформаційно-комунікаційних технологій в освіті, й зокрема, для розширення доступності освіти.

Наведені аспекти вказують на необхідність забезпечення вільного доступу до навчальної інформації і способу її представлення, всебічного висвітлення навчального матеріалу і створення умов для самоосвітньої діяльності. У цьому контексті вільне програмне забезпечення є необхідним варіантом програмного супроводу навчальної діяльності майбутніх учителів.

**Метою статті** є окреслення стратегії впровадження вільного програмного забезпечення в процес підготовки вчителів математики, фізики та інформатики з аналізом особливостей його використання.

## 2. МЕТОДИ ДОСЛІДЖЕННЯ

Для досягнення мети дослідження застосовувався комплекс методів: аналіз, систематизація, узагальнення психолого-педагогічної і методичної літератури з проблем використання вільного програмного забезпечення в освітній діяльності з метою виявлення актуальних напрямів дослідження; аналіз навчальних планів і робочих програм підготовки вчителів математики, фізики та інформатики з метою виявлення ступеня використання вільного програмного забезпечення у підготовці вчителів; аналіз передового і масового педагогічного досвіду впровадження вільного програмного забезпечення у вищих навчальних закладах.

## 3. РЕЗУЛЬТАТИ ДОСЛІДЖЕННЯ

Вільне програмне забезпечення, за формулюванням його автора Річарда Столлмана, повинно відповідати зазначеним чотирьом ступеням: свободи використання, доступністю коду, вільним розповсюдженням та видозміненням. Наразі автор програмного забезпечення зберігає за собою надані йому законом авторські права навіть за відкритості коду програмного забезпечення. Відкритий доступ до вихідних текстів програм є ключовою ознакою вільного програмного забезпечення, тому запропонований дещо пізніше Еріком Реймондом термін відкритого програмного забезпечення в деяких випадках є більш точним для позначення феномену вільного програмного забезпечення. Відмінність між вільним і відкритим програмним забезпеченням полягає в наданих пріоритетах: прихильники відкритого програмного забезпечення роблять акцент на ефективності відкритих вихідних кодів як методу розробки; прихильники вільного програмного забезпечення вважають, що безпосередньо, право на розповсюдження, модифікацію та вивчення програмного забезпечення є його головною перевагою. Для поєднання цих термінів створили спеціальну категорію Free/Libre and Open-Source Software. Абревіатура FLOSS, через слово французького походження libre, використовується переважно в структурах Європейського Союзу, а термін FOSS у США [9].

Програмне забезпечення, яке використовується в освітній діяльності, повинно відповідати санітарно-гігієнічним, навчально-методичним та функціонально технічним вимогам. Попри це слід враховувати наступні фактори у виборі програмного забезпечення [8]:
- ліцензійна «чистота» програмного забезпечення, яка передбачає не тільки використання його виключно в стінах навчального закладу, а також й на домашніх і мобільних платформах студентів і викладачів;
- зменшення фінансового навантаження на навчальний заклад і, обов'язково, на домашні і мобільні системи студентів і викладачів;
- підвищена функціональність програмного забезпечення з можливістю його «тонкого» налаштування і можливої адаптації до нових стандартів;
- стабільність використання програмного забезпечення, яке передбачає незалежність від розробника і якомога довша технічна підтримка програмного забезпечення.

На практиці комплектація програмними засобами в навчальних закладах ведеться на основі компетентностей викладачів і обслуговуючого персоналу з врахуванням існуючого методичного супроводу, який пристосований до певного програмного забезпечення, зазвичай пропрієтарного.

Під час підготовки студентів гуманітарних напрямків використання вільного програмного забезпечення мінімальне. З одного боку, у деяких напрямах використання інформаційно-комунікаційних технологій вільного програмного забезпечення не існує взагалі, або воно не відповідає вимогам, а з іншого, – тисячі пакетів і десятки тисяч програм, не відомих фахівцям як через проблему локалізації, так і через проблему розповсюдження опису програмного забезпечення.

Використання вільного програмного забезпечення у підготовці студентів фізико-математичного напрямку представлена незначною кількістю найбільш відомих програм і акетів. Так у підготовці вчителів математики, фізики та інформатики використовуються системи комп'ютерної алгебри Maxima та GAP, система математичних обчислень Octave і комп'ютерної математики Scilab, пакети статистичного аналізу R та PSPP, середовища розробки програмного забезпечення Lazarus, Geany, GNU Emacs, система підготовки тексту LaTeX і пакети офісних додатків Apache OpenOffice, LibreOffice. Тим не менш, такої кількості вільного програмного забезпечення явно недостатньо, адже тільки на основі вільного програмного забезпечення можна підготувати висококваліфікованого фахівця, який не обмежений у виборі конкретної реалізації і здатний виконати аналіз і вибрати програмний продукт, який найбільш підходить для розв'язання поставленої перед ним задачі.

Процес упровадження програмного забезпечення в навчальну діяльність розпочинається з етапу ініціації, який планово переходить до етапу реалізації і констатувального етапу. Ініціатива щодо впровадження вільного програмного забезпечення може бути локальною (ініціатива викладачів, ІТ-спеціалістів супроводу, керівників навчальних закладів) і державною. У будь-якому варіанті програмне забезпечення повинно пройти не тільки експертну оцінку, а й пілотний експеримент із впровадження, а тому, зазвичай, державна ініціатива є результатом успішної реалізації локальних ініціатив.

Експертна оцінка програмного забезпечення, яке пропонується до впровадження, повинна враховувати як вищезазначені, так і специфічні фактори. Супровід вільного програмного забезпечення достатньою мірою представлений тільки для популярних програмних продуктів, операційних систем та їх утиліт, баз даних та середовищ розробки програмного забезпечення, web-орієнтованих програм та ігор. Для іншої частини вільного програмного забезпечення підтримка полягає в наявності тематичних англомовних форумів, засобів відстеження помилок і низці інших вузькоспеціалізованих служб, а тому не йде мова про спеціалізовану підтримку, що накладає додаткові умови на вибір вільного програмного забезпечення.

Автори вільного програмного забезпечення створюють не тільки стабільне і корисне програмне забезпечення, а й виконують роль розробників інноваційних ідей. Тим не менш, великі корпорації, через можливість тримати у своєму штаті науково-дослідні центри і лабораторії, співпрацю з інститутами й університетами, не тільки створюють вільні програмні продукти, а й займаються дослідницькою й інноваційною діяльністю, яка задає напрямки розвитку програмного забезпечення. Для такої діяльності залучаються кошти з моделі непрямої монетизації бізнесу, тобто без створення програмного продукту із закритим кодом [1].

Незалежність від конкретного постачальника передбачає, що всю відповідальність за експлуатацію вільного програмного забезпечення несе користувач і в разі технічних або організаційних проблем їх розв'язання необхідно виконати самостійно, зазвичай без залучення кваліфікованої допомоги. Така ситуація складається не тільки для популярних вільних програмних продуктів. Тематичні форуми підтримки вільного програмного забезпечення, зазвичай. дуже стримані в наданні корисних порад.

Одним із результатів роботи прикладного програмного забезпечення є інформація, яка зберігається в електронних документах, що призначена для подальшої роботи. Використання закритих форматів електронних документів прив'язує користувачів до певного програмного продукту або певного розробника на відміну від відкритих форматів електронних документів, що не тільки дають свободу вибору програмного забезпечення, а й продовжують життєвий цикл створеної і набутої інформації.

Етап реалізації впровадження програмного забезпечення розпочинається з аналізу проблеми, дослідження її вирішеності й аналізу існуючих результатів. Розглядаються можливості технічної реалізації впровадження програмного забезпечення та відповідність технічних характеристик обчислювальних систем заявленим потребам. Виконується всебічний аналіз існуючих форм і методів навчання, з якими повинно узгоджуватися програмне забезпечення, а в разі невідповідності проводяться роботи з внесення змін в нього.

Упровадження вільного програмного забезпечення, через специфіку його створення, дозволяє активізувати навчально-пізнавальну діяльність під час виконання поданих нижче етапів.

1) Локальна ініціатива з упровадження вільного програмного забезпечення, яка базується на факторах необхідності й можливості його використання.
2) Аналіз існуючого досвіду використання вільного програмного забезпечення в навчальній діяльності.
3) Аналіз стримуючих факторів використання вільного програмного забезпечення і пошук рішень з кожного з факторів.

4) Пошук й експертна оцінка вільного програмного забезпечення з врахуванням можливих ризиків його використання.
5) Технічна експертиза програмного забезпечення і відповідність заявлених вимог технічним характеристикам наявної обчислювальної техніки.
6) Проведення пілотного проекту з використання вільного програмного забезпечення, розробка методичного супроводу.
7) Аналіз результатів пілотного проекту й у разі їх позитивності розповсюдження набутого досвіду.
8) Державна ініціатива з упровадження вільного програмного забезпечення, створення і розповсюдження методичного супроводу, підготовка й перепідготовка фахівців з його використання.
9) Аналіз отриманих результатів і їх публікація.

Загальні рекомендації впровадження вільного програмного забезпечення в кожному конкретному випадку потребують уточнення залежно як від самого програмного забезпечення, так і від середовища його використання. Зокрема, стосовно вчителів математики, фізики та інформатики, у підготовці яких є навчальні предмети з програмування, слід запланувати виконання курсової або кваліфікаційної роботи, головним результатом якої є створення програмного продукту навчального призначення на засадах вільного програмного забезпечення. Створені так програмні продукти мають не тільки зрозумілі функції, а й відповідають компонентам освітнього стандарту. Попри це, послідовно виконуючи подібні роботи, можна створити репозиторій необхідного для студентів і викладачів програмного забезпечення.

Іншим видом навчальної діяльності студентів є локалізація вільного програмного забезпечення. Так як вихідні коди відкриті, а в більшості програмних продуктів механізм локалізації виділено в окремий електронний документ, то можливість адаптації програмного забезпечення не тільки привнесе більше зручностей у використанні вільного програмного забезпечення, а й надасть студентам досвіду з розробки програмного забезпечення, можливість колективної/групової роботи, дотримання стандартів.

Наступним видом діяльності може бути пошук й аналіз існуючого вільного програмного забезпечення з метою його апробації і використання в науково-дослідній роботі. Такий вид навчальної діяльності необхідно супроводжувати опублікованими в мережі Інтернет звітами результатів досліджень, що призведе до популяризації знайдених програмних продуктів і методик їх використання.

Також, не менш значущим у підготовці майбутніх учителів математики, фізики та інформатики є створення ними медіа контенту навчального матеріалу з розповсюдженням його засобами хмарних технологій. Як сам процес створення, так і процес його поширення дасть змогу отримати цінний практичний дослід створення медіа супроводу навчального матеріалу.

## 4. ВИСНОВКИ ТА ПЕРСПЕКТИВИ ПОДАЛЬШИХ ДОСЛІДЖЕНЬ

Вільне програмне забезпечення показало свою «життєву здатність» в багатьох сферах: науці, державному управлінні, банківській справі, освіті. Для майбутніх учителів математики, фізики та інформатики вільне програмне забезпечення повинно бути не тільки засобом навчання, а й предметом вивчення. Базуючись на ідеях вільного програмного забезпечення і використовуючи відкриті вхідні коди програм, студенти названих спеціальностей, через суттєву інформатичну підготовку, повинні створювати нові програмні продукти і користуватись без обмежень вже існуючими. А тому аналіз можливостей використання вільного програмного забезпечення в навчальній діяльності студентів фізико-математичних спеціальностей і поступове його впровадження є нагальною проблемою вищої освіти.

Напрямки подальшого дослідження створення навчально-методичного забезпечення щодо використання вільного програмного забезпечення у підготовці майбутніх учителів фізики, математики та інформатики.

## СТРАТЕГИЯ ВНЕДРЕНИЯ СВОБОДНОГО ПРОГРАММНОГО ОБЕСПЕЧЕНИЯ В ПРОЦЕСС ПОДГОТОВКИ УЧИТЕЛЕЙ МАТЕМАТИКИ, ФИЗИКИ И ИНФОРМАТИКИ

**Величко Владислав Евгеньевич**
канд. физ-мат наук, доцент, докторант
ГУ «Луганский национальный университет имени Тараса Шевченко», г. Старобельск, Украина
*vladislav.velichko@gmail.com*

**Аннотация.** Информационные процессы в обществе побуждают образование к пересмотру форм и методов обучения, предполагающие использование широких дидактических возможностей информационно-коммуникационных технологий. Не менее значима в этом контексте проблема подготовки специалистов, которые способны использовать современные возможности компьютерной техники. Подготовка высококвалифицированного учителя возможна только при условии использования в его обучении передовых технологий, которые охватывают весь спектр существующих возможностей. Анализ используемого в образовании программного обеспечения показал недостаточное использование в образовательном процессе целого класса программного обеспечения — свободного программного обеспечения. Для преодоления этой проблемы предложена стратегия внедрения свободного программного обеспечения при подготовке учителей математики, физики и информатики.

**Ключевые слова:** свободное программное обеспечение; стратегия внедрения; подготовка учителей математики, физики и информатики.

# IMPLEMENTATION STRATEGY OF FREE SOFTWARE IN THE PROCESS OF PREPARATION OF TEACHERS OF MATHEMATICS, PHYSICS AND COMPUTER SCIENCE

**Vladyslav Ye. Velychko**
PhD (Physical and Mathematical Sciences), associate professor, doctoral candidate
Luhansk Taras Shevchenko National University, Starobilsk, Ukraine
*vladislav.velichko@gmail.com*

**Abstract.** Information processes in the society encourage the formation of a revision of the forms and methods of learning; involve the use of didactic capabilities of information and communication technologies in teaching. No less important in this context, the problem of professionals training who are able to use modern possibilities of computer technology. Training of highly qualified teachers is only possible using advanced technologies that cover the entire range of existing opportunities. The analysis used in the formation of the software has showed insufficient use of a whole class of software - free software in the educational process. To overcome this problem, the proposed implementation strategy of free software in the preparation of teachers of mathematics, physics and computer science is proposed.

**Keywords:** free software; implementation strategy; the training of teachers of mathematics, physics and computer science.